\documentclass[twocolumn,showpacs,preprintnumbers,amsmath,amssymb]{revtex4}
\pdfoutput=1
\usepackage{graphicx}
\usepackage{dcolumn}
\usepackage{bm}
\usepackage{subfigure}

\begin{document}

\title{Phases of Holographic Superconductors in an External Magnetic Field}

\author{Tameem  Albash}
 \email{talbash [at] usc.edu}
\author{Clifford V. Johnson}
 \email{johnson1 [at] usc.edu}
\affiliation{%
Physics and Astronomy Department, University of Southern California, Los Angeles, 
CA 90089--0484, USA
}%

\date{2nd June 2009}

\begin{abstract}
  We study a 2+1 dimensional model of superconductors using a 3+1
  dimensional gravitational dual theory of a black hole coupled to a
  scalar field, with negative cosmological constant. In the presence
  of finite temperature $T$ and a background magnetic field $B$, we
  use numerical and analytic techniques to solve the full
  Maxwell--scalar equations of motion in the background geometry,
  finding non--trivial localized solutions that correspond to
  condensate droplets, and to vortices. The properties of these
  solutions enable us to deduce several key features of the $(B,T$)
  phase diagram.
\end{abstract}


\maketitle

\section{\label{sec:level1}Introduction}
For slightly over a decade, there has been a growing branch of string
theory research that extracts important physics of strongly coupled
systems by computing with a weakly coupled gravitational ``dual''
system. These ``gauge/gravity'' dualities ---termed thus since the
strongly coupled system is a gauge field theory or generalization
thereof--- are said to be holographic in nature, since the dual
gravitational system crucially has at least one extra dimension, and
much of the field theory's properties can be extracted by working on
the boundary of the gravitating spacetime. The best understood version
of this is the AdS/CFT
correspondence\cite{Maldacena:1997re,Witten:1998qj,Gubser:1998bc}
between superstring theory on AdS$_5\times S^5$(4+1 dimensional
anti--de-Sitter times a five--sphere) and 3+1 dimensional ${\cal
  N}{=}4$ supersymmetric $SU(N)$ Yang--Mills theory. The theory can be
reduced on the $S^5$ to a gravity theory coupled to a family of fields
in the AdS$_5$. The boundary of AdS$_5$ is a copy of 3+1 dimensional
Minkowski space, and this is the spacetime where the dual Yang--Mills
theory resides. Its 't Hooft coupling, $\lambda{=}4\pi g_sN{=}g_{\rm
  YM}^2N$, is large in the regime where the dual gravity model is
reliable: the large~$N$ limit and weak string coupling $g_s$. The
asymptotic values of the AdS$_5$ fields supply information about
operators in the Yang--Mills theory.  Ref.\cite{Aharony:1999t} reviews
of much of this technology, with bibliography. This type of duality
has supplied a wealth of information about various strongly coupled
systems that can be supplied with a gravitational (or fully stringy)
dual along these lines. It has several potentially important
applications, ranging from models of the behaviour of the strongly
coupled dynamics of quark--gluon plasmas in nuclear physics, to models
relevant to condensed matter physics, all of considerable experimental
interest. (See
refs.\cite{Hartnoll:2009sz,Herzog:2009xv,Natsuume:2007qq} for reviews
and bibliography.)

A holographic model of some of the key phenomenological attributes of
superconductivity in 2+1 dimensions was proposed in
ref.\cite{Hartnoll:2008vx}. The dual is a simple model of gravity in four
dimensions (AdS$_4$) coupled to a $U(1)$
gauge field and a minimally coupled charged complex scalar $\Psi$ with
potential $V(|\Psi|)=-2|\Psi|^2/L^2$, where the cosmological constant
defines the scale $L$ {\it via} $\Lambda=-3/L^2$:
\begin{eqnarray} \label{eqt:EM_action}
S_{\mathrm{bulk}} &=& \frac{1}{2 \kappa_4^2}  \int d^4 x \sqrt{-G} \biggl\{ R + 
\frac{6}{L^2} + \\ &&L^2 \left(-\frac{1}{4} F^2 - \left| \partial \Psi - i g A \Psi \right|^2 - V \left(\left|\Psi \right| \right)\right) \biggr\}\ , \nonumber
\end{eqnarray}
where $\kappa_4^2=8\pi G_{\rm N}$ is the gravitational coupling and
our signature is $(-+++)$. We will use coordinates $(t, z, r,\phi)$
for much of our discussion, with $t$ time, $(r,\phi)$ forming a plane,
and $z$ a ``radial'' coordinate for our asymptotically AdS$_4$
spacetimes such that $z=0$ is the boundary at infinity.  Note that the
mass of the scalar $m^2_\Psi=-2/L^2$ is above the
Breitenlohner--Freedman stability bound\cite{Breitenlohner:1982bm}
$m^2_{\rm BF}=-9/4L^2$ for scalars in AdS$_4$.  A black hole (which is
planar, its horizon is an $(r,\phi)$ plane at some finite $z=z_h$),
with Hawking temperature $T$ and mass per unit horizon area
$\varepsilon=M/V$, corresponds to the dual 2+1 dimensional system at
temperature $T$ and with energy density~$\varepsilon$.  The asymptotic
value of $\Psi$ on the boundary sets the vacuum expectation value
(vev) of a charged operator~$\cal O$ in the system, which is the order
parameter. For $T>T_c$, the scalar (and hence $\langle{\cal
  O}\rangle$) is zero, and for $T<T_c$ it is non--zero. In the
gravitational theory, the high $T$ phase is simply the charged black
hole (Reissner--Nordstr\"{o}m, or AdS--RN) with $\Psi$ vanishing. Notice
that the mass of the scalar is set not just by $V(|\Psi|)$ but by the
black hole's gauge field $A=A_tdt$. ($A_t$ does not give an electric
field in the dual theory on $(r,\phi)$, but defines instead a $U(1)$
charge density\cite{Chamblin:1999tk}, $\rho$.  See below.) $T$ also
depends on $\rho$. In fact $m^2_\Psi$ decreases with $T$ until
at~$T_c$ it goes below $m^2_{\rm BF}$, becoming tachyonic. The theory
seeks a new solution, in which the black hole is no longer 
AdS--RN, but one that has a non--trivial profile for~$\Psi$ ({\it
  i.e.}, it has ``scalar hair'' --- for a discussion of violations of
non--hair theorems in this context, see ref.\cite{Gubser:2008px}). The
$U(1)$ is broken by $\langle{\cal O}\rangle$.  These solutions can be
found by solving equations of motion  in certain limits (explored
below), and the transport properties of the low temperature phase were
examined in refs.\cite{Hartnoll:2008vx,Hartnoll:2008kx}, using linear
response theory, with the result that the DC conductivity diverges in
a way consistent with expectations that the phase is
superconducting. (Strictly speaking, the $U(1)$ that was broken by
$\langle{\cal O}\rangle$ is global on the boundary, but it can be
gauged in a number of ways without affecting the conclusions. See
{\it e.g.} ref.\cite{Hartnoll:2008kx}.)

It is clearly of interest to study this system further, since it could
well open the door to a whole new phenomenology of superconductivity
in a wide range of physical systems. We report here on our study of
the system in an external magnetic field, continuing the work we began
in ref.\cite{Albash:2008eh}.  The magnetic field $B$ (which fills the
two dimensions of the superconducting theory), also contributes to
$m^2_{\Psi}$, {\it via} its square, but contributes with {\it opposite
  sign} to the electric contribution of the background.  It therefore
lowers the temperature $T_c$ at which $m^2_\Psi$ falls below $m^2_{\rm
  BF}$, triggering the phase transition.  On these grounds alone one
then expects a critical line in the $(B,T)$ plane of a form like that
of a full curve from fig.~\ref{fig:pd_unite}, but it is important to
determine exactly what physics lies on either side of the
line. Generically, for non--zero $B$, it is inconsistent to have
non--trivial spatially independent solutions on the boundary, and we
study two classes of localized solutions in some detail. The first is
a ``droplet'' solution, the prototype of which was found in our
earlier work\cite{Albash:2008eh} as a strip in 2D (straightforwardly
generalized to circular symmetry in ref.\cite{Hartnoll:2008kx}), and
the second is a vortex solution, with integer winding number $\xi\in
{\cal Z}$, which is entirely new.  We obtain these as full solutions
of the Maxwell--scalar sector in a limit, and their properties allow
us to determine key features of the $(B,T)$ phase diagram (correcting
statements made in ref.\cite{Hartnoll:2008kx}).

\section{\label{independent}Spatially Independent Solutions}
This section reviews the spatially independent condensate solution of
ref.\cite{Hartnoll:2008vx} that corresponds to the superconducting (symmetry
breaking) phase below some $T_c$. It is found in a certain decoupling
limit. Under a redefinition $A_\mu\to A_\mu/g$, $\Psi\to\Psi/g$, the
Maxwell--scalar part of the action~(\ref{eqt:EM_action}) gets a
prefactor of $1/g^2$, and the $g$ disappears from the $A\Psi$
coupling. This means that in the limit $g\to\infty$, this sector
decouples from gravity. We can therefore take as background
black hole the (planar) Schwarzschild solution:
\begin{equation}\label{schwarzschild}
ds^2=\frac{L^2\alpha^2}{z^2}\left(-f(z)dt^2+dr^2+r^2d\phi^2\right)+\frac{L^2}{z^2f(z)^2}dz^2\ , 
\end{equation}
with $f(z){=}1-z^3$. It has an horizon at $z{=}1$, temperature
$T{=}3\alpha/4\pi$, and mass density
$\varepsilon{=}L^2\alpha^3/\kappa^2_4$. Let us write our complex scalar
as $\Psi{=}{\tilde\rho}\exp(i\theta)/\sqrt{2}L$ and write $A_t{=}\alpha
{\tilde A}t$. The equations of motion allow an ansatz: $\theta{=}{\rm
  const.}$, ${\tilde\rho}{=}{\tilde\rho}(z)$, ${\tilde A}_t{=}{\tilde
  A}_t(z)$, and $A_\phi{=}0$. Near the boundary $z{=}0$ we have, for constants
${\tilde\rho}_1,{\tilde\rho}_2,\rho,\mu$:
\begin{equation}
{\tilde\rho}\to{\tilde\rho}_1 z+{\tilde\rho}_2 z^2\ , \quad {\tilde A}_t\to \mu-\rho z\ .
\end{equation}
${\tilde\rho}_i$ ($i{=}1,2$) sets the vev of a $\Delta{=}i$ operator
${\cal O}_i$ \cite{Klebanov:1999tb}. Only one of these vevs can be
non--zero at a time, and we will choose to study the case of $i{=}1$,
for brevity. $\mu$ sets a chemical potential, and $\rho$ sets a charge
density for the $U(1)$ symmetry that we are studying here. To find a
full solution of the equations, we solve the equations of motion
numerically, using a shooting method starting at the horizon
($z{=}1$). There we set ${\tilde\rho}(1)$ to a constant, and for
regularity of the gauge field, we have ${\tilde A}_t(1){=}0$. To look
for solutions we tune $\partial_z{\tilde A}_t(1)$ such that at the
other boundary, ${\tilde\rho}(0)$ has the required Neumann condition
${\tilde\rho}_2{=}0.$ From that we read off ${\tilde\rho}_1{=}\sqrt{2}
\kappa_4\langle{\cal O}_1\rangle/(L\alpha)$, and~$\rho$. Since the
only scale in the theory besides $T$ is set by the charge density
$\rho$, the value of $T_c$ is given in terms of~$\rho$. This can be
determined by noting when a non--zero $\langle{\cal O}_1\rangle$
develops, and there we find $T_c{=}0.226\alpha\sqrt{\rho}$.
Fig.~1 
shows the result for the vev as a function of temperature, showing the
low and high temperature phases separated by a second order phase
transition at $T_c$ below which the $U(1)$ is spontaneously broken.
\begin{figure}[h]
\includegraphics[width=2.0in]{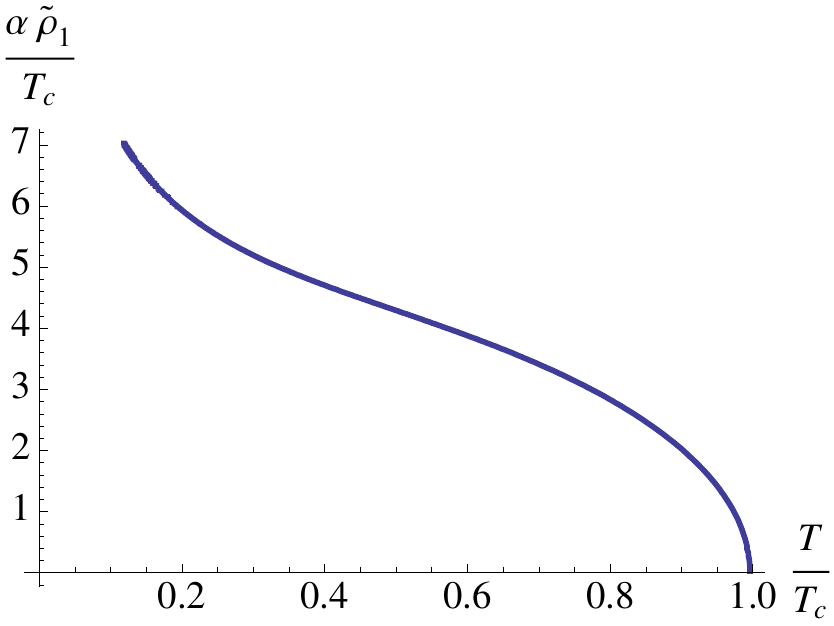}\label{fig:constant_sol_rho1}
\caption{The scalar ${\tilde \rho}\sim\langle{\cal O}_1\rangle$ {\it vs.} $T$. $T_c=0.226\alpha\sqrt{\rho}$.}
\end{figure}
\section{Spatially Dependent Solutions}
We can seek non--trivial solutions with $\theta{=}\zeta+\xi\phi$,  and
\begin{equation}
{\tilde\rho}={\tilde\rho}({\tilde r},z), {\tilde A}_t={\tilde A}_t({\tilde r},z), A_\phi={\tilde A}_\phi({\tilde r},z)\ ,
\end{equation}
where ${\tilde r}{=}\alpha r$ and $(\zeta,\xi)$ are constants, with
$\xi$ integer. Regularity of the equations of motion require that near
${\tilde r}{=}0$, we must have ${\tilde\rho}{\sim} {\tilde
  r}^\xi$. This feeds into the behaviour of all the other fields near
${\tilde r}{=}0$, and determining this near the horizon is important
in order to seed the numerical search in a manner that will carefully
find solutions.  We report on this in a longer
publication\cite{longer}. The solutions that we find using our
analysis break into two broad classes. There are those for which
${\tilde\rho}\to0$ for ${\tilde r}\to\infty$ which are called
droplets. The prototype localized solution of this type was found in
our earlier work\cite{Albash:2008eh}, and also studied in
ref.\cite{Hartnoll:2008kx}, but in a ``probe'' limit of small fields
and finite $g$. Here we have the full solutions in the large $g$
decoupling limit. Those solutions for which ${\tilde\rho}\to {\rm
  const.}$ for ${\tilde r}\to\infty$ we call vortex solutions
($\xi\in{\cal Z}^+$). They behave exactly as expected of vortices in
this context. Note that~$\xi$ defines a non--trivial topological
winding number. ${\tilde A}_\phi$ becomes constant at infinity so
gauge symmetry cannot be used to unwind $\theta$. Gauge symmetry is
unbroken at infinity for the droplets, so $\xi$ is not a winding
number for them.
\subsection{\label{droplets} The Droplet}
\begin{figure}[h]
\subfigure[Scalar Field for $\gamma =0.5$]{\includegraphics[width=2.5in]{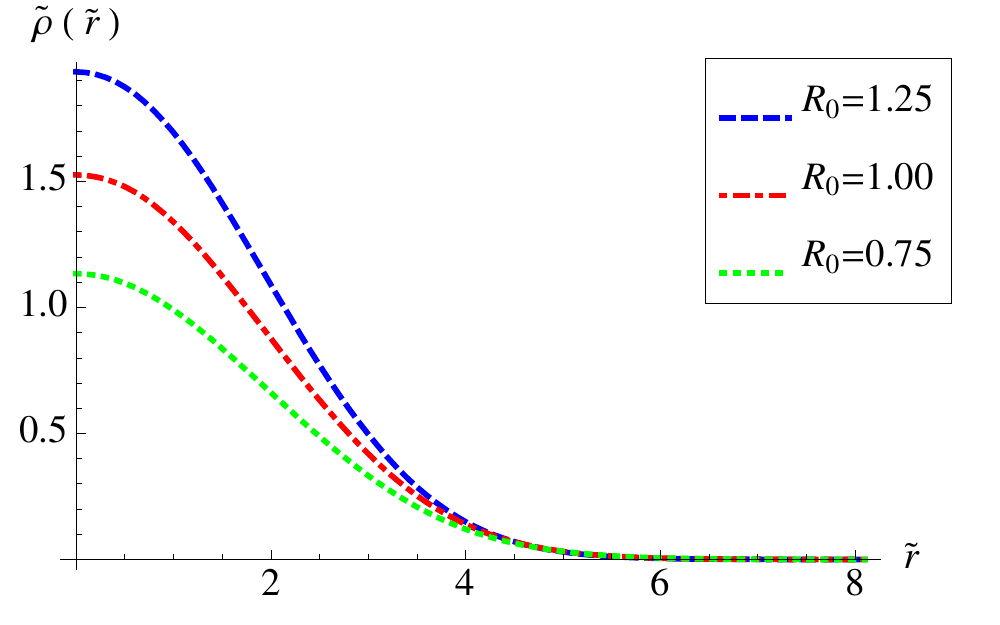}\label{fig:droplet_rho_gamma=0.5}} 
\subfigure[Magnetic Field for $\gamma =0.5$]{\includegraphics[width=2.5in]{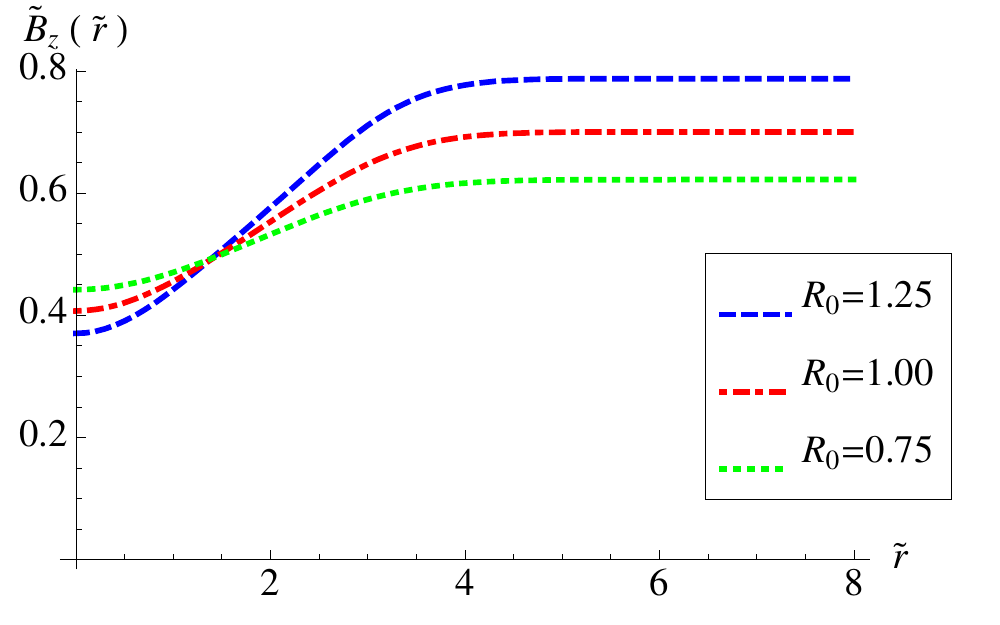}\label{fig:droplet_B_gamma=0.5}}
\caption{\small A droplet solution for $\gamma =0.5$ at  $T/T_c \approx 0.84$.} \label{fig:droplet_solution}
\end{figure}

A sample droplet solution is presented in
fig.~\ref{fig:droplet_solution} (for~${\tilde\rho}$ and ${\tilde
  B}_z$). The current density and magnetic field are read off {\it
  via} the $z\to0$ asymptotic ${\tilde A}_\phi\to
a_\phi+J_\phi({\tilde r})z$, with ${\tilde B}_z{=}\partial_{{\tilde
    r}}a_\phi/{{\tilde r}}$.
The constant $\gamma$ appears in the small~${\tilde r}$ expansion of
the function ${\tilde R}$ defined by ${\tilde \rho}=z{\tilde
  r}^\xi{\tilde R}$, where ${\tilde R}\to R_0\left(1-\gamma{\tilde
    r}^2/4+\cdots\right)$ at the horizon.  We see that the magnetic
field fills the plane, asymptoting to a constant value at ${\tilde
  r}\to\infty$, and approaching a non--zero value in the core. We
studied several solutions for a range of $T$ and $B\equiv B_z$,
noticing that for high temperatures $B$ is decreased somewhat in the
core as compared to the asymptotic value, while for low temperatures
$B$ is enhanced there.

\begin{figure}[h] 
   \centering
   \includegraphics[width=2.6in]{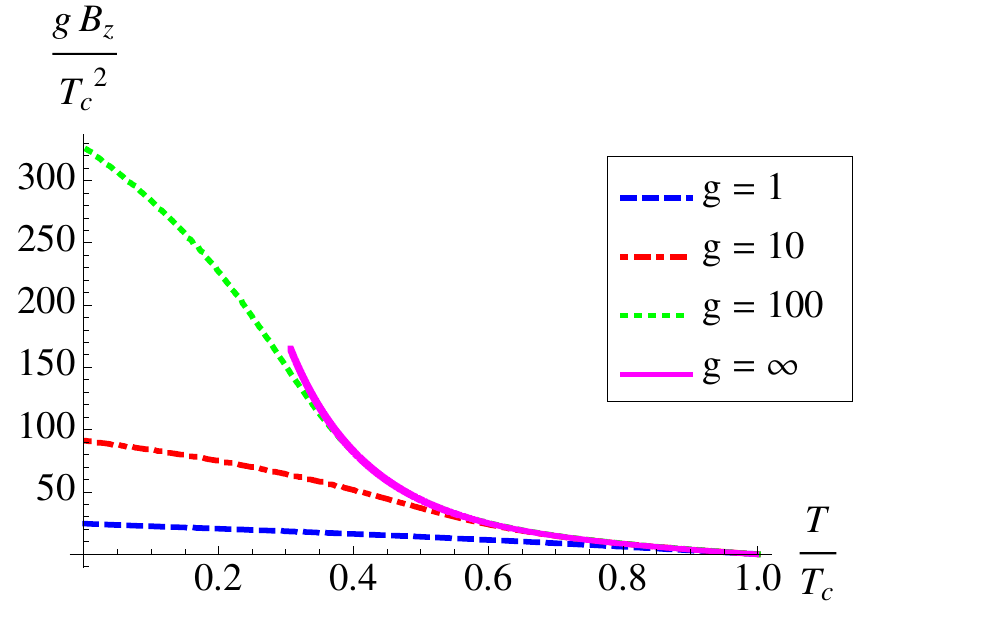} 
   \caption{\small Scaled $(B,T)$ diagram  connecting to
     decoupling limit.}
   \label{fig:pd_unite}
\end{figure}

It is important to determine exactly where in the $(B,T)$ plane these
solutions can appear. We find our full solutions at a variety of
values of $B$, but none below certain values of
$B$, which is suggestive. To determine if there is a minimum value of
$B$ to form droplets, we can use the probe limit, where ${\tilde
  \rho}$ is small. Here we can take $\tilde{A}_\phi=\gamma{\tilde
  r}^2/2$ (which will determine the magnetic field in this limit) and
have $\tilde{A}_t={\tilde A}_t(z)$. The equation of motion for
$\tilde{R}$ on the horizon reduces to:
\begin{equation}\label{hermite}
\left[\partial_{\tilde{r}}^2 \tilde{R} + \frac{1}{\tilde{r}} \partial_{\tilde{r}} \tilde{R} -\frac{1}{4} \tilde{r}^2 {\gamma}^2 \tilde{R} + \gamma \tilde{R} \right]_{z=1}  = 0\ ,
\end{equation}
and has as solution: $\tilde{R}(\tilde{r}) {=} R_0(1) \exp
\left(-\gamma {\tilde{r}^2}/{4} \right)$. We use this as the seed to
solve the coupled equations for ${\tilde A}_t$ and ${\tilde R}$ for
the full $z$ dependence in this limit, seeking the appropriate
boundary conditions at $z{=}0$.  This extracts, as in
section~\ref{independent}, the $T$ of the solution for our magnetic
field.  This procedure sets the critical magnetic field at which the
droplet solutions first form for a given $T/T_c$. Below that $B$ the
droplets simply vanish since in this probe limit they are already at
zero size.  Our result is the top curve in fig.~\ref{fig:pd_unite}.
Large fields will begin to back--react on the geometry, and our
analysis breaks down, as can be seen in our curve for low $T/T_c$. To
proceed further, we can carry out the probe computation again, but not
in the $g\to\infty$ limit. Since the Maxwell sector is not decoupled
from the geometry for finite $g$ we use as background a dyonic black
hole. This is the analysis of our earlier
work\cite{Albash:2008eh}. The metric is of the form given in
eqn.~(\ref{schwarzschild}), but with a Maxwell field $F= 2 h \alpha^2
r dr \wedge d \phi + 2 q \alpha dz \wedge dt$ and $f(z)=\left(1-z
\right)\left( z^2 +z+ 1 - \left(h^2 +q^2\right) z^3 \right)$. Now, $T$
and~$\rho$ are determined by the background (we do not display them
here). We note that the equations of motion have a separable solution
${\tilde\rho}=zZ(z)R({\tilde r})$, and $R({\tilde r})$ satisfies an
equation of the same form as~(\ref{hermite}), but with $\gamma\to2gh$,
giving the Gaussian profile. We can then solve for $Z(z)$ numerically
using the same shooting techniques as before. This gives a complete
$(B,T)$ curve for a given~$g$.  In fact, these curves connect to the
$g\to\infty$ probe computation of above, as can be established by
careful comparison of the different definitions of the temperatures
and charge densities in the two limits. We carry this out in our
longer paper\cite{longer}, and simply show here how the two limits
connect by rescaling our curves (with the appropriate factor of $g$)
and superposing. See fig.~\ref{fig:pd_unite}.

\subsection{\label{vortices}The Vortex}
\begin{figure}[h!]
\begin{center}
\subfigure[Scalar Field]{\includegraphics[width=2.5in]{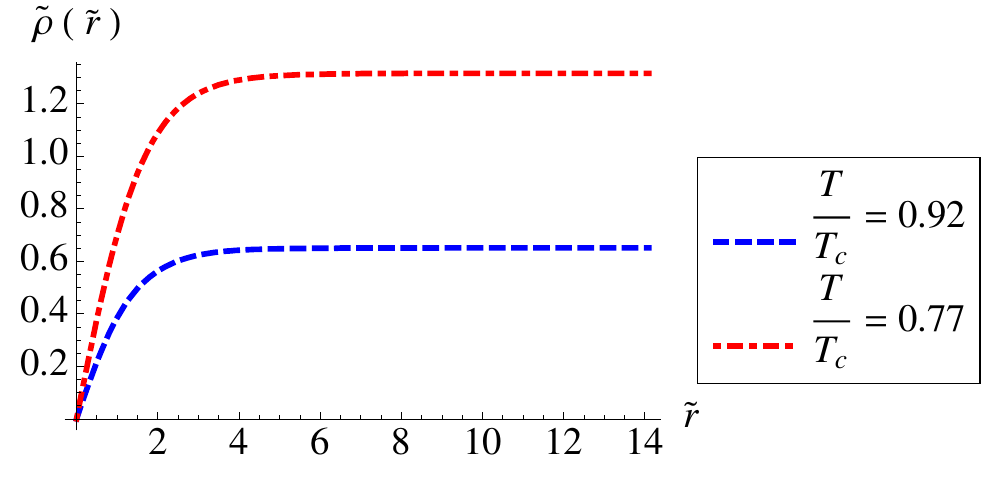}\label{fig:vortex_linear_rho1}}
\subfigure[Magnetic Field]{\includegraphics[width=2.5in]{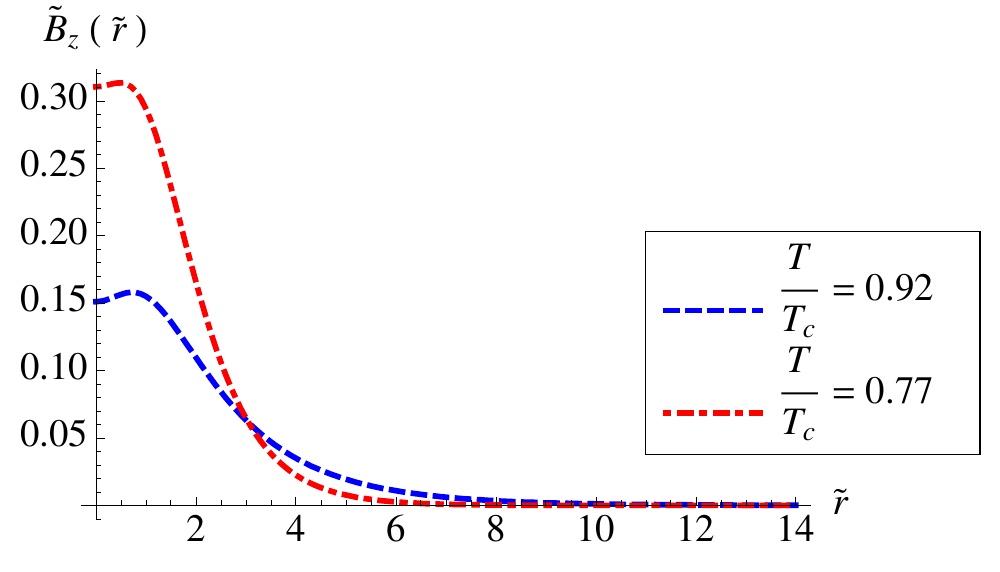}\label{fig:vortex_linear_magnetic}}
\caption{\small The vortex solution for $\xi = 1$.} \label{fig:vortex_solution}
\end{center}
\end{figure}

We display the $\xi=1$ vortex
solution 
in fig.~\ref{fig:vortex_solution} (for ${\tilde\rho}$ and~${\tilde
  B}_z$). As noted before, there is one unit of winding, and the
scalar field runs to a constant at infinity. So does the charge
density, and so we are able to read off the value of the temperature
for these solutions. The magnetic field drops to zero at large
${\tilde r}$, and ${\tilde A}_\phi$ becomes constant there. That
constant is $\xi$ in general, as is consistent with the total magnetic
flux being quantized to $2\pi\xi$. The current density $J_\phi({\tilde
  r})$ (not displayed for brevity) is zero asymptotically and peaks in
a ring around the core, supporting the magnetic field, as expected for
a vortex.
As for the droplets, the question arises as to where in the $(B,T)$
plane vortices appear. As mentioned, we determine $T/T_c$ from the
large~${\tilde r}$ values of $\rho$ and ${\tilde
  \rho}|_{z=0}=\sqrt{2}\kappa_4\langle{\cal O}_1\rangle/(L\alpha)$,
while the magnetic field is localized in the core. For an applied $B$
at some $T/T_c$, we expect that the system creates vortices to trap
the flux in their cores, forming a lattice of them that increases in
density with $B$.  Vortices presumably repel each other, and so such a
lattice will cost energy. Therefore at some critical $B_c(T)$ the
system will seek a lower energy phase, possibly returning to the
normal phase. We do not know if this critical line coincides with the
one in fig.~\ref{fig:pd_unite} where the droplets begin, although it
is the simplest possibility.

\section{Conclusion}
We have studied the prototype holographic superconductor of
ref.\cite{Hartnoll:2008vx} in the presence of magnetic field,
continuing our earlier work\cite{Albash:2008eh}, examining two
important classes of solution, the droplet and the vortex. The latter
is completely new, and this is the first time the droplet has been
fully constructed. Our analysis in various connected limits shows
where these solutions can exist in the $(B,T)$ plane. There is a
critical line below which droplets are not found, while vortices can
be found there. Our interpretation is that this region is the
superconducting phase, and that for non--zero $B$, the vortices
develop, trapping the magnetic flux into filaments, as is familiar in
type~II superconductors. Above the critical line, the system leaves
the superconducting phase, and either forms droplets of condensate or
simply reverts to the normal phase (dual to a dyonic black hole with
zero scalar everywhere, which may well yield lower action than the
droplets if we had back--reacting solutions to work with).  We have
performed a stability analysis of these solutions\cite{longer}. Both
classes are stable for the examples and modes that we study (this is
expected for the vortices given their conserved winding). Finally,
note that we disagree with the suggestion made in
ref.\cite{Hartnoll:2008kx}. The authors find the critical line, but
state (similarly to our ref.\cite{Albash:2008eh}) that the droplets
exist {\it below} the line, and are superconducting. As they did not
have the full droplet solutions, nor the vortex solutions, their
analyses are not sufficient to make these determinations. Our work
here gives a much stronger picture of the phase diagram.

\begin{acknowledgments}
  This work was supported by the Department of Energy. We thank Arnab
  Kundu and Rob Myers for useful conversations.
\end{acknowledgments}


\begin{thebibliography}{15}
\expandafter\ifx\csname natexlab\endcsname\relax\def\natexlab#1{#1}\fi
\expandafter\ifx\csname bibnamefont\endcsname\relax
  \def\bibnamefont#1{#1}\fi
\expandafter\ifx\csname bibfnamefont\endcsname\relax
  \def\bibfnamefont#1{#1}\fi
\expandafter\ifx\csname citenamefont\endcsname\relax
  \def\citenamefont#1{#1}\fi
\expandafter\ifx\csname url\endcsname\relax
  \def\url#1{\texttt{#1}}\fi
\expandafter\ifx\csname urlprefix\endcsname\relax\def\urlprefix{URL }\fi
\providecommand{\bibinfo}[2]{#2}
\providecommand{\eprint}[2][]{\url{#2}}

\bibitem[{\citenamefont{Maldacena}(1998)}]{Maldacena:1997re}
\bibinfo{author}{\bibfnamefont{J.~M.} \bibnamefont{Maldacena}},
  \bibinfo{journal}{Adv. Theor. Math. Phys.} \textbf{\bibinfo{volume}{2}},
  \bibinfo{pages}{231} (\bibinfo{year}{1998}), \eprint{hep-th/9711200}.

\bibitem[{\citenamefont{Witten}(1998)}]{Witten:1998qj}
\bibinfo{author}{\bibfnamefont{E.}~\bibnamefont{Witten}},
  \bibinfo{journal}{Adv. Theor. Math. Phys.} \textbf{\bibinfo{volume}{2}},
  \bibinfo{pages}{253} (\bibinfo{year}{1998}), \eprint{hep-th/9802150}.

\bibitem[{\citenamefont{Gubser et~al.}(1998)\citenamefont{Gubser, Klebanov, and
  Polyakov}}]{Gubser:1998bc}
\bibinfo{author}{\bibfnamefont{S.~S.} \bibnamefont{Gubser}},
  \bibinfo{author}{\bibfnamefont{I.~R.} \bibnamefont{Klebanov}},
  \bibnamefont{and} \bibinfo{author}{\bibfnamefont{A.~M.}
  \bibnamefont{Polyakov}}, \bibinfo{journal}{Phys. Lett.}
  \textbf{\bibinfo{volume}{B428}}, \bibinfo{pages}{105} (\bibinfo{year}{1998}),
  \eprint{hep-th/9802109}.

\bibitem[{\citenamefont{Aharony et~al.}(2000)\citenamefont{Aharony, Gubser,
  Maldacena, Ooguri, and Oz}}]{Aharony:1999t}
\bibinfo{author}{\bibfnamefont{O.}~\bibnamefont{Aharony}},
  \bibinfo{author}{\bibfnamefont{S.~S.} \bibnamefont{Gubser}},
  \bibinfo{author}{\bibfnamefont{J.~M.} \bibnamefont{Maldacena}},
  \bibinfo{author}{\bibfnamefont{H.}~\bibnamefont{Ooguri}}, \bibnamefont{and}
  \bibinfo{author}{\bibfnamefont{Y.}~\bibnamefont{Oz}}, \bibinfo{journal}{Phys.
  Rept.} \textbf{\bibinfo{volume}{323}}, \bibinfo{pages}{183}
  (\bibinfo{year}{2000}), \eprint{hep-th/9905111}.

\bibitem[{\citenamefont{Hartnoll}(2009)}]{Hartnoll:2009sz}
\bibinfo{author}{\bibfnamefont{S.~A.} \bibnamefont{Hartnoll}}
  (\bibinfo{year}{2009}), \eprint{0903.3246}.

\bibitem[{\citenamefont{Herzog}(2009)}]{Herzog:2009xv}
\bibinfo{author}{\bibfnamefont{C.~P.} \bibnamefont{Herzog}}
  (\bibinfo{year}{2009}), \eprint{0904.1975}.

\bibitem[{\citenamefont{Natsuume}(2007)}]{Natsuume:2007qq}
\bibinfo{author}{\bibfnamefont{M.}~\bibnamefont{Natsuume}}
  (\bibinfo{year}{2007}), \eprint{hep-ph/0701201}.

\bibitem[{\citenamefont{Hartnoll
  et~al.}(2008{\natexlab{a}})\citenamefont{Hartnoll, Herzog, and
  Horowitz}}]{Hartnoll:2008vx}
\bibinfo{author}{\bibfnamefont{S.~A.} \bibnamefont{Hartnoll}},
  \bibinfo{author}{\bibfnamefont{C.~P.} \bibnamefont{Herzog}},
  \bibnamefont{and} \bibinfo{author}{\bibfnamefont{G.~T.}
  \bibnamefont{Horowitz}}, \bibinfo{journal}{Phys. Rev. Lett.}
  \textbf{\bibinfo{volume}{101}}, \bibinfo{pages}{031601}
  (\bibinfo{year}{2008}{\natexlab{a}}), \eprint{0803.3295}.

\bibitem[{\citenamefont{Breitenlohner and
  Freedman}(1982)}]{Breitenlohner:1982bm}
\bibinfo{author}{\bibfnamefont{P.}~\bibnamefont{Breitenlohner}}
  \bibnamefont{and} \bibinfo{author}{\bibfnamefont{D.~Z.}
  \bibnamefont{Freedman}}, \bibinfo{journal}{Phys. Lett.}
  \textbf{\bibinfo{volume}{B115}}, \bibinfo{pages}{197} (\bibinfo{year}{1982}).

\bibitem[{\citenamefont{Chamblin et~al.}(1999)\citenamefont{Chamblin, Emparan,
  Johnson, and Myers}}]{Chamblin:1999tk}
\bibinfo{author}{\bibfnamefont{A.}~\bibnamefont{Chamblin}},
  \bibinfo{author}{\bibfnamefont{R.}~\bibnamefont{Emparan}},
  \bibinfo{author}{\bibfnamefont{C.~V.} \bibnamefont{Johnson}},
  \bibnamefont{and} \bibinfo{author}{\bibfnamefont{R.~C.} \bibnamefont{Myers}},
  \bibinfo{journal}{Phys. Rev.} \textbf{\bibinfo{volume}{D60}},
  \bibinfo{pages}{064018} (\bibinfo{year}{1999}), \eprint{hep-th/9902170}.

\bibitem[{\citenamefont{Gubser}(2008)}]{Gubser:2008px}
\bibinfo{author}{\bibfnamefont{S.~S.} \bibnamefont{Gubser}}
  (\bibinfo{year}{2008}), \eprint{0801.2977}.

\bibitem[{\citenamefont{Hartnoll
  et~al.}(2008{\natexlab{b}})\citenamefont{Hartnoll, Herzog, and
  Horowitz}}]{Hartnoll:2008kx}
\bibinfo{author}{\bibfnamefont{S.~A.} \bibnamefont{Hartnoll}},
  \bibinfo{author}{\bibfnamefont{C.~P.} \bibnamefont{Herzog}},
  \bibnamefont{and} \bibinfo{author}{\bibfnamefont{G.~T.}
  \bibnamefont{Horowitz}} (\bibinfo{year}{2008}{\natexlab{b}}),
  \eprint{0810.1563}.

\bibitem[{\citenamefont{Albash and Johnson}(2008)}]{Albash:2008eh}
\bibinfo{author}{\bibfnamefont{T.}~\bibnamefont{Albash}} \bibnamefont{and}
  \bibinfo{author}{\bibfnamefont{C.~V.} \bibnamefont{Johnson}},
  \bibinfo{journal}{JHEP} \textbf{\bibinfo{volume}{09}}, \bibinfo{pages}{121}
  (\bibinfo{year}{2008}), \eprint{0804.3466}.

\bibitem[{\citenamefont{Klebanov and Witten}(1999)}]{Klebanov:1999tb}
\bibinfo{author}{\bibfnamefont{I.~R.} \bibnamefont{Klebanov}} \bibnamefont{and}
  \bibinfo{author}{\bibfnamefont{E.}~\bibnamefont{Witten}},
  \bibinfo{journal}{Nucl. Phys.} \textbf{\bibinfo{volume}{B556}},
  \bibinfo{pages}{89} (\bibinfo{year}{1999}), \eprint{hep-th/9905104}.

\bibitem[{\citenamefont{Albash and Johnson}(2009)}]{longer}
\bibinfo{author}{\bibfnamefont{T.}~\bibnamefont{Albash}} \bibnamefont{and}
  \bibinfo{author}{\bibfnamefont{C.~V.} \bibnamefont{Johnson}}
  (\bibinfo{year}{2009}), \eprint{{\it to appear.}}

\end{thebibliography}

\end{document}